\documentclass[useAMS,usenatbib]{mn2e}
\usepackage{graphicx}
\usepackage{lscape}
\usepackage{times}
\usepackage{amsmath}
\usepackage{amsfonts}
\usepackage{amssymb}
\usepackage{multirow}
\usepackage{longtable}
\usepackage{afterpage}
\RequirePackage{color}

\title[Fundamental properties from frequencies]{
Fundamental properties of solar-like oscillating stars from frequencies 
of minimum  $\Delta \nu$: II. Model computations 
{for different chemical compositions and mass}
}
\author[M. Y\i ld\i z, Z. \c{C}elik Orhan and C. Kayhan]{M. Y\i ld\i z$^{}$\thanks{E-mail:
mutlu.yildiz@ege.edu.tr}, Z. \c{C}elik Orhan and C. Kayhan \\
Department of Astronomy and Space Sciences, Science Faculty, Ege University, 35100, Bornova, \.Izmir, Turkey.}

\begin{document}
%\onecolumn
\date{Accepted 2013 May 15. Received 2013 April 11; in original form 2013 April 11}

\pagerange{\pageref{firstpage}--\pageref{lastpage}} \pubyear{2013}
\def\braket#1{\left<#1\right>}
\newcommand{\wrt}{\mbox{with respect to }}
\newcommand{\numin}{\mbox{\ifmmode{\nu_{\rm min}}\else$\nu_{\rm min}$\fi}}
\newcommand{\teff}{\mbox{\ifmmode{T_{\rm eff}}\else$T_{\rm eff}$\fi}}
\newcommand{\teffsun}{\mbox{\ifmmode{{\rm T}_{\rm eff\sun}}\else${\rm T}_{\rm eff\sun}$\fi}}
\newcommand{\numax}{\mbox{$\nu_{\rm max}$}}
\newcommand{\Dnu}{\mbox{$\Delta \nu$}}
\newcommand{\muHz}{\mbox{$\mu$Hz}}
\newcommand{\kepler}{\mbox{\it{Kepler}}}
\newcommand{\corot}{\mbox{\it{CoRoT}}}
\newcommand{\numaxS}{\mbox{$\nu_{\rm max \sun}$}}
\newcommand{\MS}{{\rm M}\ifmmode_{\sun}\else$_{\sun}$~\fi}
\newcommand{\RS}{{\rm R}\ifmmode_{\sun}\else$_{\sun}$~\fi}
\newcommand{\LS}{{\rm L}\ifmmode_{\sun}\else$_{\sun}$~\fi}
\newcommand{\MSbit}{{\rm M}\ifmmode_{\sun}\else$_{\sun}$\fi}
\newcommand{\RSbit}{{\rm R}\ifmmode_{\sun}\else$_{\sun}$\fi}
\newcommand{\LSbit}{{\rm L}\ifmmode_{\sun}\else$_{\sun}$\fi}

\maketitle

\label{firstpage}

\begin{abstract}
The large separations between the oscillation frequencies 
of solar-like stars are measures of stellar mean density. 
The separations have been thought to be mostly constant in 
the observed range of frequencies. 
However, detailed investigation shows that they are not constant, and their
variations are not random but have very strong diagnostic potential 
for our understanding of stellar structure and evolution.
In this regard, frequencies of the minimum large separation
are very useful tools.
From these frequencies, in addition to the large separation and frequency 
of maximum amplitude, Y\i ld\i z et al. recently 
have developed new methods to 
find almost all the fundamental stellar properties.
In the present study, we aim to find metallicity and helium abundances 
from the frequencies, and generalize the relations given by 
Y\i ld\i z et al. for a wider stellar mass range and arbitrary 
metallicity ($Z$) and helium abundance ($Y$).
We show that the effect of metallicity is { significant}
for most of the fundamental parameters.
For stellar mass, for example, the expression must be multiplied by 
$(Z/Z_{\sun})^{0.12}$.
For arbitrary helium abundance, $M\propto (Y/Y_{\sun})^{0.25}$.
Methods for determination of $Z$ and $Y$ from pure asteroseismic 
quantities are based on amplitudes (differences between maximum and 
minimum values of \Dnu) in the oscillatory component in the spacing 
of oscillation frequencies. 
Additionally, we demonstrate that the difference between 
the first maximum and the second minimum is very sensitive to $Z$.
It also depends on $\numin_1/\numax$ and small separation between 
the frequencies. 
Such a dependence leads us to develop a method to find 
$Z$ (and $Y$) from oscillation frequencies.
The maximum difference between the estimated and model $Z$ values is 
about 14 per cent. 
It is 10 per cent for $Y$.

\end{abstract}


\section{Introduction}
Determination of fundamental stellar properties is required in many sub-fields of astrophysics.
In this regard, the promise of asteroseismology is very important, in particular for solar-like 
oscillating stars. {Most of the oscillations, at least for main-sequence (MS) stars,}  are acoustic pressure waves and their frequencies 
depend on sound speed throughout stellar interior. Sound speed, however, depends on the first adiabatic exponent
$\Gamma_1$, which  is very low in the He {\scriptsize II}  ionization zone. 
The He {\scriptsize II}  ionization zone of the solar-like
oscillating stars is not too shallow, 
%The first adiabatic exponent $\Gamma_1$ is significantly low 
and its upper and lower borders coincide with
the nodes of certain modes. It has so significant effect on the oscillation frequencies 
that one can infer basic properties of such stars from these frequencies. The present study develops 
new methods for this purpose.

%\textcolor{blue}{SORRY, A PROBLEM HERE.  YOU REALLY NEED TO START THE PAPER
%WITH A GENERAL SET OF STATEMENTS THAT WILL TELL ME WHAT YOU ARE TALKING
%ABOUT (PULSATIONS OF SOLAR STARS), AND WHY I MIGHT BE INTERESTED IN THEM.
%THEN IN THE 2ND PARAGRAPH GO AHEAD AND TALK ABOUT MATH/PHYSICS RELATIONSHIPS.
%ALSO, I THINK YOU ONLY USE THE WORKD EIGENFREQUENCIES HERE AND ONE OTHER
%TIME IN THE PAPER.  MAYBE DROP THE TERM COMPLETELY (I DON'T RELLAY KNOW WHAT
%YOU ARE TRYING TO SAY IN THIS SENTENCE), AND GO RIGHT TO THE
%PHYSICS OF STARS?}

%The eigenfrequencies of any system exhibit its main characteristic features. 
{Solar-like oscillating stars have such regular oscillation 
frequencies ($\nu$) that the frequency of a mode linearly depends on its 
order $n$ and degree $l$.} % : $\nu_{nl}=(n+l/2+\alpha_n)$) 
This dependence is known as the asymptotic relation. 
{According to this idea, the frequencies of modes with 
adjacent orders and the same $l$ are evenly spaced by the so-called 
large separation, $\Delta \nu= \nu_{nl}-\nu_{n-1,l}$. 
In reality, however, there are some deviations from this simple relation,
and these variations lead us to discover parameters related to stellar 
structure and evolution.}
For this purpose, Y{\i}ld{\i}z et al. (2014; 
hereafter Paper I)
introduce two new reference frequencies $\numin_1$ and $\numin_2$, which are the frequencies at which the 
large separations between the oscillation frequencies are minimum. 
In Paper I, {new expressions} for fundamental stellar properties, such as stellar mass ($M$), radius ($R$), surface gravity ($g$),  luminosity ($L$), effective temperature ($T_{\rm eff}$) and age ($t$), are derived from 
the interior models of 0.8-1.3 \MS with solar composition.  
The present study aims to generalize these expressions {for the wider mass range ($1.0$-$1.6$ \MS) 
than this range} and to test the effects of the metallicity ($Z$) and helium abundance ($Y$)
on these relations. We also  try to determine
$Y$  and  $Z$ %if chemical composition can be found 
from the oscillation frequencies,
at least for interior models.
%new expressions for chemical composition in terms of oscillation frequencies. 

{ The frequencies of MS models are used in deriving 
expressions for age and other quantities.
For sub-giants and giants, these relations must be tested.
}

This paper is organized as follows. 
{In Section 2, we make general considerations about the reference frequencies.}
% and show how oscillatory component in \Dnu $ $ is shaped by 
%%the first adiabatic exponent (
%$\Gamma_1$ 
%in the He {\scriptsize II} ionization zone.
Section 3 is devoted to {generalizing} the relations obtained in Paper I for wider mass range and arbitrary
metallicity and helium abundance.
%determine effects of Y and Z on the scaling relations.
We present the sensitivity of the adiabatic oscillation frequencies to the
metallicity and helium abundance, and new methods for determination of $Y$ and $Z$ from oscillation frequencies 
in Section 4. 
%Section 3 is devoted to determine effects of X and Z on the scaling relations.
%We apply our methods on the Kepler and CoRoT stars and find their fundamental properties in Section 4.
Finally, in Section 5, we draw our conclusions.

%\textcolor{blue}{I THINK THAT YOU MIGHT END THE INTRODUCTION HERE.  THAT
%IS, I SUGGEST THE YOU MOVE THE PARAGRAPH ``This paper is organized ...''
%HERE, AND START A NEW SECTION, CALLED SOMETHING LIKE ``General Considerations''
%OR SOMETHING SIMILAR, WITH THE TEXT OF THIS NEXT PARAGRAPH}
%
\section{General Considerations}
{The asteroseismic parameters that can be extracted from 
oscillation frequencies and related to the stellar parameters are
\Dnu, the frequency of the maximum amplitude (\numax), and small separation between the oscillation frequencies 
($\delta \nu_{02}=\nu_{nl}-\nu_{n-1,l+2}$).}
 %Kjeldsen \& Bedding (1995, hereafter KB95) 
Brown et al. (1991) give \numax$~$as (see also Kjeldsen \& Bedding  1995)
\begin{equation}
\numax\propto \frac{M}{R^2}T_{\rm eff}^{-1/2}.
%\label{equ:M}
\end{equation}
In addition to these parameters, we introduce \numin$_1$  and \numin$_2$. 
Their ratio to \numax$~$gives us the stellar mass:
\begin{equation}
{\nu_{\rm max}}= 1.188 \nu_{\rm min1} \frac{\MS}{M}
               = 1.623 \nu_{\rm min2} \frac{\MS}{M}
%\frac{M}{\MS}=1.188\frac{\nu_{\rm min1}}{\nu_{\rm max}}
%             =1.623\frac{\nu_{\rm min2}}{\nu_{\rm max}}
\end{equation}
where the numeric values 1.188 and 1.623 come from the ratio $\numax_{\sun}/\nu_{{\rm min1} \sun}$ and $\numax_{\sun}/\nu_{{\rm min2} \sun}$, respectively.
%\textcolor{blue}{WHERE DO THESE NUMERIC VALUES COME FROM?}
If we insert equation (2) in equation (1), expression for $\nu_{\rm min1}$ and $\nu_{\rm min2}$ in terms of
fundamental stellar parameters is derived as 
\begin{equation}
\numin_1\propto \numin_2\propto \frac{M^2}{R^2}T_{\rm eff}^{-1/2}.
%\label{equ:M}
\end{equation}
{Equation (3) is valid at least for the models with solar values and mass ranging from 0.8 \MS to 1.3 \MS presented 
in Paper I.}

Stellar parameters change with chemical composition. 
%{ Increases in $Z$ and $Y$, for example, cause stellar
%radius $R$ to increase.}
{ Therefore, one can expect that $\numax$ is 
also a sensitive function of $Z$ and $Y$,  and thus equation (1) will need to be modified to take 
these effects into account. }
{For the relations derived in Paper I, it is also important 
to understand how the other asteroseismic quantities are influenced by 
change in $Z$ and $Y$ (see Section 3.2).}

%{\bf
%In stellar astrophysics, we try to explain basic features of stars in 
%terms of their masses. $R$ and $L$ of MS stars, for example, are in great extent determined by $M$.
%However, there is no single $M$-$R$ or $M$-$L$ relation for the entire mass range we deal.
%There is a transition mass ($M_{\rm t}$) such that
%two different relations are valid for $M>M_{\rm t}$ and $M<M_{\rm t}$.
%$M_{\rm t}$ may also be considered as the transition mass from late-type to 
%early-type stars and is found by Y\i ld\i z et al. (2014b) as 1.30 \MS for the solar composition. %dikkat 1.45 olacak
%For high metallicity, however, this transition mass becomes very high.
%A star with $Z=0.04$ and $M=1.75$ \MSbit, for example, may in some respects 
%be considered as a late-type star. 
%Such a star may have solar-like oscillating features. 
%Thus, chemical composition, in particular metallicity, is important as 
%much as stellar mass in many cases.}

{Inference of the solar helium surface abundance ($Y_{\rm s}$) 
from high degree ($l> 40$) oscillation frequencies of the Sun is the subject 
of many studies.} %The result is very precise. 
Basu \& Antia (1995), for example, give $Y_{\rm s}$
as $0.25 \pm 0.01$.
Unfortunately, high degree oscillations are not observable for other stars.  
Houdek \& Gough (2007) confirm that the amplitude of the second difference
of the frequencies 
depends on the helium abundance. 
%\textcolor{blue}{SECOND DIFFERENCE OF WHAT?  THIS IS UNCLEAR}
Recently, {Verma et al. (2014) find helium abundances from $Kepler$ data for 16 Cyg A and B.
In Paper I, we have used the frequencies at which \Dnu $ $ 
is minimum for determination of stellar parameters.
In the present study, however, we use the difference between the maximum 
and minimum values of $\Dnu$ for determination of helium abundance (see Section 4.1). }
A similar approach may also lead us to develop a new method for 
inference of metallicity from oscillation frequencies.

\section{General Relations for Stellar Parameters from asteroseismic quantities}
{ 
In this study, as in Paper I, the stellar interior models with solar chemical composition 
are constructed 
by using the {\small ANK\.I} code (Ezer \& Cameron 1965; Y{\i}ld{\i}z 2011).
The solar chemical composition is taken as $X = 0.7024$, $Y=0.2804$
and $Z = 0.0172$.
The adiabatic oscillation frequencies of these models are computed by 
{\small ADIPLS} (Christensen-Dalsgaard 2008). } 
{In this section, we test if the relations derived in Paper I are also 
valid for $1.3$ \MS $< M \leq 1.6$ \MS and arbitrary $Z$ and $Y$.}

{Both of the minima do not appear in all the models within the mass range { we deal with}. 
For example, the second minimum is
not seen in the oscillation frequencies of the models with $M < 1.0$ \MS (see table 1 of Paper I)
while the first minima disappears in some
models within the upper mass range of 1.3-1.6 \MSbit. Therefore, our present analysis is based on the frequency of 
the second minimum of the models with mass range 1.0-1.6 \MSbit.}

\subsection{Expressions for stellar parameters for the mass range $1.0$-$1.6$ \MS}
In Paper I, the expressions for stellar mass in terms of \numax $ $ and one 
of \numin$_1$ or \numin$_2$ are derived from the interior models for 
the mass range 0.8-1.3 \MSbit. 
%\textcolor{blue}{YOU PROBABLY SHOULD QUOTE THESE EQUATIONS HERE SINCE YOU
%DERIVE A DIFFERENT ONE BELOW FOR HIGHER MASS STARS}
%Both $\numin_1$ and  $\numin_2$ are available for these models. 
For the Sun, \numax$_{\sun} = 3050$ \muHz $ $ is higher than both 
$\numin_{1\sun}=2555.18$ $\muHz $ and  $\numin_{2\sun}=1879.52$ $\muHz$.
For 1.2 \MS models with solar composition \numax $ $ is equal to $\numin_1$. 
For the models with $1.2$ \MS $< M < 1.45$ \MSbit, $\numin_2<\numax<\numin_1$. 
And, $\numin_2>\numax$ for the models with $M>1.45$ \MS (see {Fig. 2}).
%{ As well known, \numax $ $ is the oscillation with the maximum amplitude, found by applying 
%a gaussian fit to the  power spectrum. According to this fit,
%the frequency of a mode closer to  \numax$ $ is, the higher the amplitude is.}
Oscillation amplitudes of these models with frequencies around $\numin_2$ 
are greater than that of around \numin$_1$.
{$M_2$ is the mass computed from $\nu_{\rm min2}$ (see equation 10 of Paper I),
$M_2/\MS=1.623 \numin_2/\numax $.
%The masses computed from equation (10) in terms of $\nu_{\rm min1}$ ($M_1$) and $\nu_{\rm min2}$ ($M_2$)
In {Fig. 1}, $M_2$ with $X_{\rm c}=0.17$, 0.35, 0.53 and 0.7}
is plotted with respect to model mass. 
%for the mass range 1.0-1.6 \MSbit. 
The model mass and mass found from the ratio of frequencies are in very 
good agreement for $M<1.3$ \MSbit. 
For higher-mass models, however, a deviation occurs.
Here, we obtain a relation between mass and the frequency ratio 
\numin$_2$/\numax\ {for the range $M=1.3$-$1.6$ \MS\ of}
\begin{equation}
%\frac{M}{\MS}=(0.315 \frac{\numin_2}{\numax}-0.263)^{0.603}+1.30.
\frac{M}{\MS}=(0.462 \frac{\numin_2}{\numax}-0.356)^{0.74}+1.254.
%\label{equ:M}
\end{equation}
{We note that the mass still is found from the same 
frequency ratio $\numin_2/\numax$.}
The reason for the deviation of mass from the expression given in 
equation 10 of Paper I is due to change in properties of \numin$_2$ for 
relatively higher masses.
{In {Fig. 2}, \numin$_2$, \numin$_1$ and \numax$~$ of models are plotted with respect 
to model mass.
\numax $ $ gradually decreases from 2600 to 1000 \muHz $ $ as 
model mass increases.
\numin$_2$ and \numin$_1$,} however, { decrease} in the low-mass range (1.0-1.4 \MSbit) and 
{ increase} in the high-mass range (1.4-1.6 \MSbit). 
{There is a minimum for \numin$_2$ of about 1.4 \MSbit.}
Therefore, the ratio \numin$_2$/\numax $ $ does not directly give mass.
If $M>1.3$ \MSbit, the method for computation of stellar mass can be still 
based on this ratio.  
First, we can compute stellar mass from equation 10 of Paper I. 
If the mass is greater than 1.3 \MS, then we use {equation (4)}.
 The maximum difference between the mass computed from {equation (4)} and model mass
for $M>1.3$ \MS is less than 0.025 \MSbit, as in Paper I.
{More realistic error analysis than this can be made in terms of uncertainties in observed frequencies 
and will be given in our third paper of this series.}

%The relation between stellar mass and asteroseismic quantities (\numin$_2$ 
%and \numax) is determined by details of the He {\scriptsize II} ionization zone 
%structure.
%Two parameters about the structure are very important. 
%They are thickness of the zone and position of the point where $\Gamma_1$ 
%is minimum. 
%The latter can be taken as the position of the zone.
%For the models with $M< 1.3 $ \MSbit, %the relative radius of 
%the zone ($r_{\rm HeII}/R_\star$) is localized around 0.98 (see Table 1) 
%and nearly remains constant. 
%It gradually increases to 0.99 as stellar mass approaches to$ $ $1.6$ \MSbit. 
%More importantly, the thickness of the zone is a very sensitive function of 
%the stellar mass. 
%The zone thickness of the models with $M< 1.3$ \MS is much greater than that of the models with $M> 1.3$ \MSbit.
%{ %It is well known that 
%As the order of mode increases}, the radial nodes come closer.
%The mode of minimum \Dnu $ $ is the mode which have two adjacent nodes 
%closest to the upper and lower borders of the He {\scriptsize II} ionization zone.
%As the zone becomes narrow the order of minimum \Dnu $ $ increases. 
%{Therefore, an increase in $n$ is an increase in frequency.}

\begin{figure}
\includegraphics[width=100mm,angle=0]{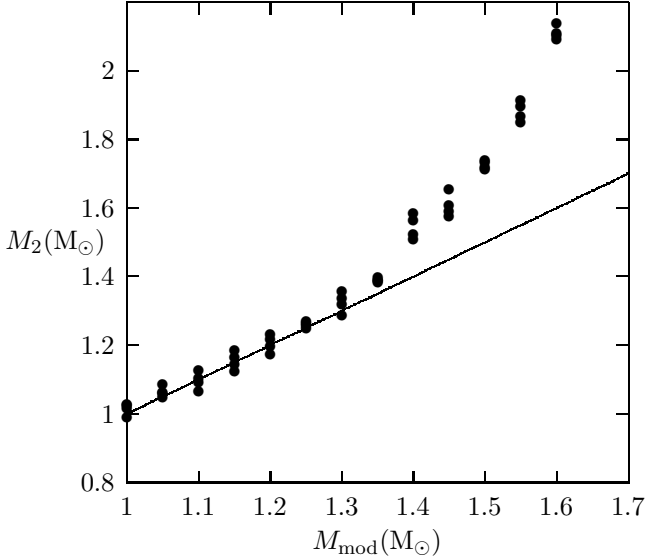}
%480,width=140mm,angle=270]{the_four_diagrams16.ps}
\caption{$M_2/\MS=1.623 \numin_2/\numax $ with respect to model mass for $X_{\rm c}=0.17$, 0.35, 0.53 and 0.7.
}
\end{figure}
\begin{figure}
\includegraphics[width=100mm,angle=0]{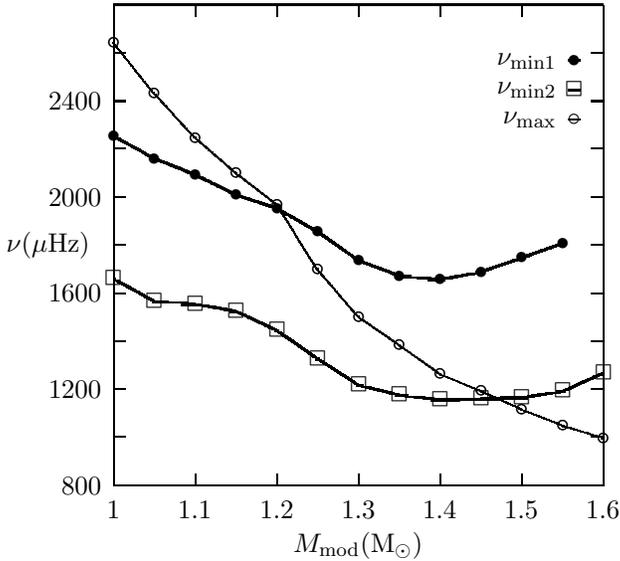}
%480,width=140mm,angle=270]{the_four_diagrams16.ps}
\caption{
{
$\nu_{\rm max}$ (circles), $\nu_{\rm min1}$ (filled circles) and $\nu_{\rm min2}$ (boxes) with respect to model mass, 
 for $X_{\rm c}=0.17$}.
}
\end{figure}
\begin{figure}
\includegraphics[width=100mm,angle=0]{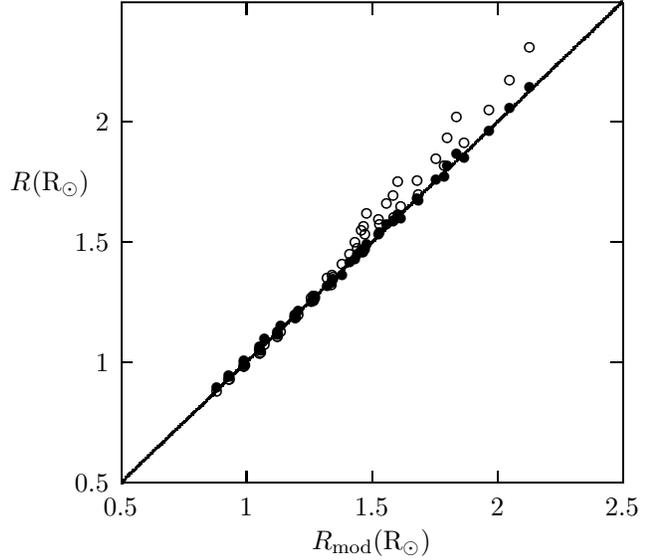}
\caption{Asteroseismic $R$ with respect to model radius.
The {open circles show the radii computed from equation 17 of Paper I and
the filled circles are } for the radii from {equation (5)}.
The radii from these equations are very close but {equation (5)} is
in better agreement with the radii of models with $M> 1.3$ \MSbit.
}
\end{figure}
\begin{figure}
\includegraphics[width=100mm,angle=0]{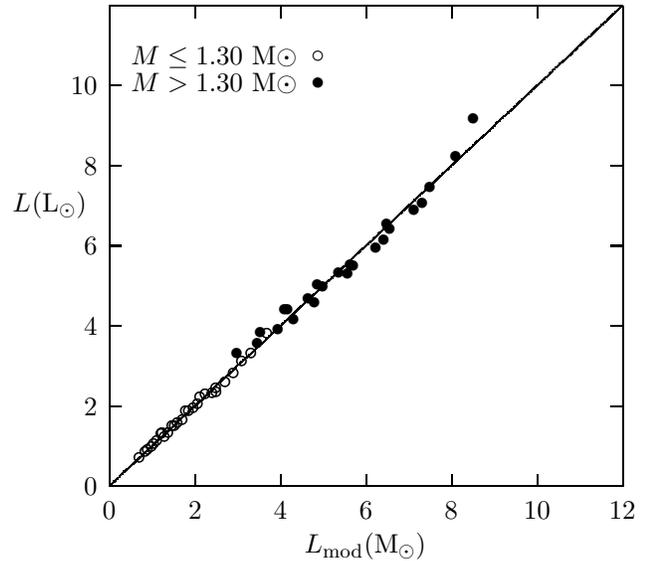}
\caption{The asteroseismic luminosity with respect to model luminosity.
Equation 20 of Paper I  and {equation (6)} are used for  the models {with $M\leq 1.30$ \MS (circles) and $M>1.30$ \MS (filled circles),
respectively.}
}
\end{figure}

%0.3152 numin2/numax-0.2627)**0.6027)+1.30

The expressions for other stellar parameters given in Paper I must also 
be tested for the mass {range 1.3-1.6}  \MSbit. 
For radius, we obtain 
\begin{equation}
%\frac{R}{{\rm R}_{\sun}}=1.009\left(\frac{\numin_2}{\nu_{\rm min2\sun}}\right)^{0.13}\left(\frac{\braket{\Delta \nu_{\sun}}}{\braket{\Dnu}}\right)^{0.87}+0.01
\frac{R}{{\rm R}_{\sun}}=1.024\left(\frac{\numin_2}{\nu_{\rm min2\sun}}\right)^{0.09}\left(\frac{\braket{\Delta \nu_{\sun}}}{\braket{\Dnu}}\right)^{0.83},
%$12/2493.2)**0.14/($2/135.5)**0.875)*1.015
\label{equ:Rmin}
\end{equation}
where $\braket{\Delta \nu}$ is the mean of $\Delta \nu$.
In {Fig. 3}, radii computed by using equation 17 of Paper I and {equation (5)}
%estimated by using \numin$_2~$ and \Dnu $ $  (equation 14 in Paper I) is 
are plotted with respect to model radii. The radii computed from {equation (5)} are in very good agreement with model radii.
{ The difference between these two radii is mostly  
less than 2 per cent.}
For the models with $M>1.3$ \MSbit, a slight difference appears
between model radius and radius from equation 17 of Paper I.

{We want to check if expressions for $L$ and $t$ for the mass range 0.8-1.3 \MS, given in equations 
20 and 22 of Paper I, respectively, are also valid for the range 1.3-1.6 \MSbit.}
{We have obtained an expression for luminosity for the 
entire range 1.0-1.6 \MSbit, but that expression has very high {departure from the model values}
for the lower part of the range, about 20 per cent.} 
Therefore, we derive a separate expression for the range 1.3-1.6 \MSbit:
%The new expression is slightly different from equation () of Paper I:
\begin{equation}
\frac{L}{\LS}=2.016\left(\frac{\nu_{\rm min2}}{\nu_{{\rm min2}\sun}}\right)^{1.3}\frac{\braket{\Delta \nu_{\sun}}}{\braket{\Dnu}}\frac{\nu_{{\rm max} \sun}}{\numax}-0.456.
%'out.numin.z172.M=1.35-1.60.Xc=ALL.fit.kontrol.pp' u 7:((2.016*(1./($2/135.15)*(($13/1879.52)**1.3*3050./(($1/$6**2/($8/5777.)**0.5*3050.))))-0.456)) lt 3 pt 7 t '    $M > 1.30$ M$\odot$'
%\frac{L}{\LS}=1.67\frac{\nu_{\rm min1}}{\nu_{{\rm min1}\sun}}\frac{\braket{\Delta \nu_{\sun}}}{\braket{\Dnu}}\frac{\nu_{{\rm max} \sun}}{\numax}-0.01.
%p 'out.numin.z172' u ($7):((1.874*($13/1887.1)**1/($2/135.15)**1)/($11/3050.)-0.729) w p pt 6 lt 3 not,x not,'out.numin.z172.Mgt1.30' u ($7):((1.874*($13/1887.1)**1/($2/135.15)**1)/($11/3050.)-0.729) w p pt 7 lt 3 not
\end{equation}
The agreement between the estimated and model luminosities is shown in {Fig. 4}.
For the estimated luminosity, equation 20 of Paper I is used {if 
$M\leq 1.3$ \MS and} {equation (6)} is employed if $M> 1.3$ \MSbit.
The difference between the estimated and model luminosities is mostly less than {8 per cent.}
We do not derive a separate fitting curve for \teff$ $ since \teff$ $ can 
easily be obtained from ${ L}$ and ${ R}$. 
For alternative expressions, see {equations (12) and (13)}.

{Equation 22 for age in Paper I is valid for masses up 
to 1.30 \MSbit.}
%For the models with $M> 1.35 \MS$, 
The estimated ages for the models with 1.35-1.6 \MS start to deviate  from the 
model ages. 
For the mass range $M> 1.30$ \MSbit, we derive expression for age as
\begin{equation}
t({\rm Gyr})=\frac{4.79\left(1.16-\frac{\braket{\delta \nu_{02}}}{\braket{\delta \nu_{02\sun}}}\right)}
{\left(\frac{\numin_2\numax_{\sun}}{\numin_{2\sun}\numax}\right)^{2.7}},
\end{equation}
%\begin{equation}
%t({\rm Gyr})=6.93r_\nu \left(0.695-0.55\frac{\braket{\delta \nu_{02}}}{\braket{\delta \nu_{02\sun}}}\right)
%  \left(3.25 - 1.54\frac{M}{{\rm M}_{\sun}} \right)
%\nonumber
%\end{equation}
%{ where 
%$$ r_\nu=\left(\frac{\numax}{\nu_{\rm max\odot}} 
%\frac{\braket{\Dnu}}{\braket{\Delta \nu_\odot}}\right)^{0.7}.
%$$
%%$r_\nu$ is as given in equation (21) of Paper I 
where  $\braket{\delta \nu_{02}}$ is the mean of $\delta \nu_{02}$.
%small separation between the frequencies ($\delta \nu_{02}$).
In {Fig. 5}, the ages computed from equation 22 of Paper I and {equation (7)} 
are plotted with respect to model ages.
%The agreement is good in particular for $M> 1.30$ \MSbit.
{ In Paper I, the difference between the estimated and model ages 
is given as less than 0.5 Gyr. In the present study, the maximum difference between the age from {equation (7)} 
and model age for the range $M>1.30$ \MS is about
0.4 Gyr.
}

\begin{figure}
\includegraphics[width=100mm,angle=0]{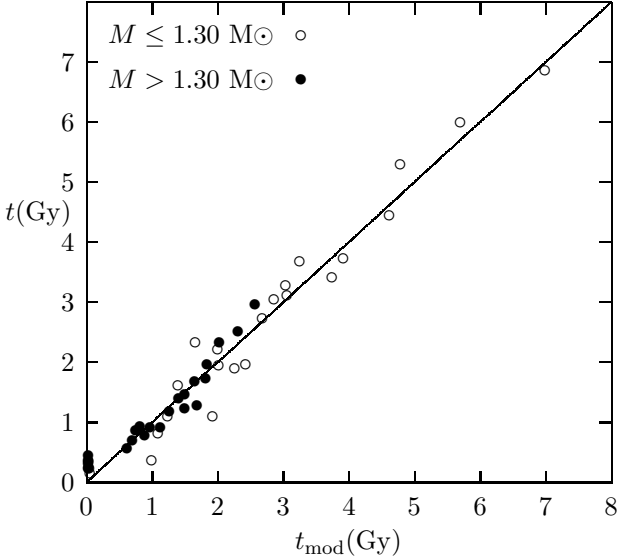}
\caption{Age inferred from asteroseismic quantities is plotted with respect to model age.
For the asteroseismic age of models {with $M\leq 1.30$ (circles) and $M> 1.30$ (filled circles),} equation 
22 of Paper I and {equation (7)} are used,
respectively.
}
\end{figure}

\subsection{Effects of chemical composition}
In Paper I, we have used solar chemical composition in the construction of 
the stellar interior models and the relations between asteroseismic and 
non-asteroseismic quantities are derived from these models. 
{In this paper, stellar models with a variety of chemical 
compositions are employed. {The mass range of these models is 1.0-1.3 \MSbit.}
The ranges of initial metallicity and helium abundance are 0.0172--0.0322 and 
0.2404--0.3204, respectively.

\subsubsection{Effects of metallicity}
Metallicity strongly influences structure and evolution of stars, and 
consequently asteroseismic properties.
Luminosity among the non-asteroseismic quantities is the most influenced 
parameter by $Z$ and  $Y$.
{It decreases as metallicity increases, while radius slightly 
diminishes, in the range we consider.}
This implies that effective temperature also decreases. 
Then, we confirm from equation (1) that 
there is a direct relation between \numax $ $ and $Z$.
The effect of metallicity on the relation  for stellar mass can be given as 
\begin{equation}
\frac{M}{\MS}=\frac{\nu_{\rm min2}}{\nu_{\rm min2\sun}}\frac{\nu_{\rm max\sun}}{\nu_{\rm max}} 
\left( \frac{Z}{{\rm Z}_{\sun}}\right)^{0.12}.
\end{equation}
%%\textcolor{blue}{IS THIS NEWLY DERIVED HERE?  IT IS NOT CLEAR}
The same equation also holds for $\nu_{\rm min1}$. % represents any of $\nu_{\rm min1}$ and $\nu_{\rm min2}$.
{{Equation (8)} is more precise than equation 23 in Paper I, 
in which the power of $(Z/{\rm Z}_{\sun})$ is found to be roughly ${0.1}$. 
These equations are derived for the first time.
%This expression 
%We obtained in Paper I (equation 6) can be given as
The $Z$ dependence in {equation (8)} is not large.
For a metal rich star, say $Z=2 {\rm Z}_{\sun}$, the effect is 
about 9 per cent. 
However, for stellar mass such an effect is very significant.}
 
Radii computed by using {equation (5)} are plotted with respect to model 
radii in {Fig. 6}.
{The slope slightly changes with $Z$: it decreases as 
$Z$ increases.}
{Equation (5)} for radius for solar metallicity is generalized for 
arbitrary $Z$ as  
\begin{figure}
\includegraphics[width=100mm,angle=0]{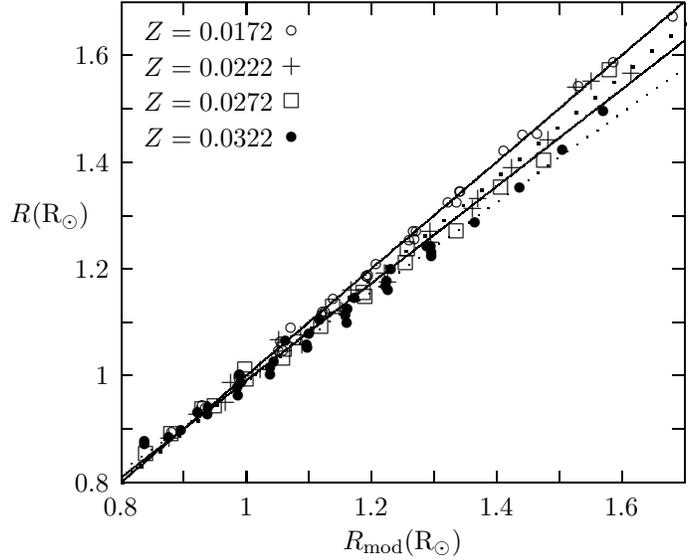}
\caption{ $R_{\rm}$ computed from {equation (5)} is plotted with respect to $R_{\rm mod}$ for different metallicities.
The upper solid line is {the fitted line} for $Z=0.0172$ and the lower dotted line is for $Z=0.0322$. 
The other lines between them are for the intermediate values of $Z$.
}
\end{figure}
\begin{figure}
\includegraphics[width=100mm,angle=0]{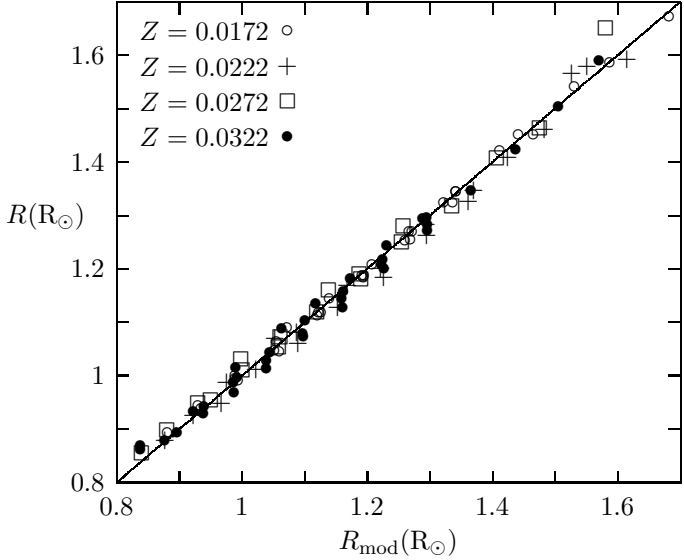}
\caption{ $R_{\rm}$ computed from {equation (9)} is plotted with respect to $R_{\rm mod}$ for different metallicities.
}
\end{figure}
\begin{equation}
\frac{R}{{\rm R}_{\sun}}=a(Z)\left(\frac{\numin_2}{\nu_{\rm min2\sun}}\right)^{0.13}\left(\frac{\braket{\Delta \nu_{\sun}}}{\braket{\Dnu}}\right)^{0.87} + b(Z),
%$12/2493.2)**0.14/($2/135.5)**0.875)*1.015
\label{equ:Rmin}
\end{equation}
where 
\begin{equation}
a(Z)=0.210\left( \frac{Z}{{\rm Z}_{\sun}}\right)+0.783
\end{equation}
and
\begin{equation}
b(Z)=-0.019\left( \frac{Z}{{\rm Z}_{\sun}}\right)^{3.32}+0.011.
\end{equation}
Radii computed from {equation (9)} are plotted with respect to model radii 
in {Fig. 7}.
The agreement is very good for the full range of $Z=0.0172$-$0.0322$. 

%R=a(ratio_min2)^0.13(ratio_Dnu)^0.87+b
%a=1/aa, b=-bb/aa
%
%a(Z)=0.210(Z/Zgun)+0.783
%b(Z)=-0.019(Z/Zgun)^3.32+0.011
%
%Modifications in equations if needed.

\begin{figure}
\includegraphics[width=100mm,angle=0]{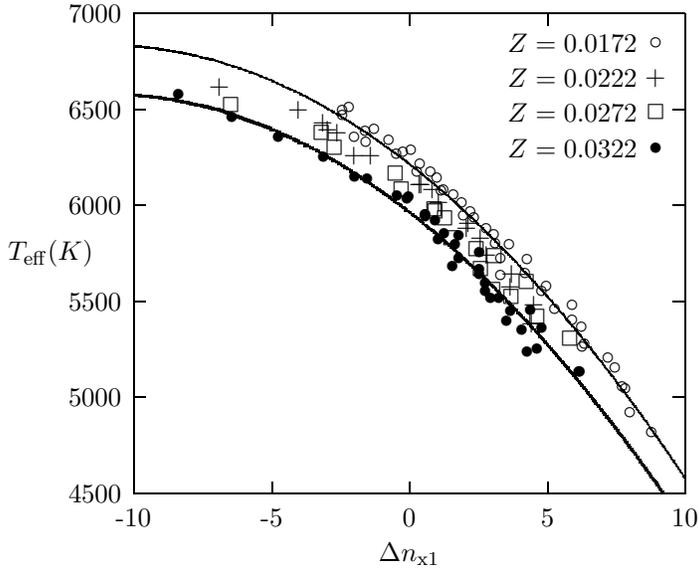}
%480,width=140mm,angle=270]{the_four_diagrams16.ps}
\caption{ $T_{\rm eff}$ with respect to $\Delta n_{\rm x1}$ for different metallicities.
The thin and thick solid lines are the fitting curves for $Z=0.0172$ and $Z=0.0322$, respectively.
}
\end{figure}

One of the very important relations obtained in Paper I is the one between 
\teff$ $ and $\Delta n_{\rm x1}$,
defined as $(\numax-\numin_1)/\Dnu$ (see fig. 6 in Paper I). 
We consider if this relation varies with $Z$.
In {Fig. 8}, \teff$ $ is plotted with respect to  $\Delta n_{\rm x1}$ for 
the models with the metallicities $Z=0.0172, ~ 0.0222,~ 0.0272$ and $ 0.0322$. 
The relation is significantly influenced by the metallicity. 
Effective temperatures of two models with the same $\Delta n_{\rm x1}$ 
but different $Z$ are different. 
The difference for the models
with  $Z=0.0172$ and $Z=0.0322$, for example,  is about 250 K.

We obtain expression for \teff$ $ as a function of $Z$ and 
$\Delta n_{\rm x1}$ as
\begin{equation}
\frac{T_{\rm eff}(Z,\Delta n_{\rm x1})}{\rm T_{\rm eff \sun}}=1.232-0.05\left( \frac{Z}{{\rm Z}_{\sun}}\right)-8.80~10^{-4}(\Delta n_{\rm x1}+11)^{2}.
\end{equation}
{ The maximum difference between \teff$ $ from {equation (12)} and model \teff$ $
is mostly less than 100 K.}
Similarly, we derive an alternative expression for \teff$ $ in terms of 
$\Delta n_{\rm x2}$, defined as 
$(\numax-\numin_2)/\Dnu$:
\begin{equation}
\frac{T_{\rm eff}(Z,\Delta n_{\rm x2})}{\rm T_{\rm eff \sun}}=1.275-0.061\left( \frac{Z}{{\rm Z}_{\sun}}\right)-6.06~10^{-4}(\Delta n_{\rm x2}+6)^{2.2}.
%p -3.5*(x+6)**2.2+(-0.061*(0.0272/0.0172)+1.275)*5780
\end{equation}

These two expressions ({equations 12 and 13}) for \teff$ $ are very important 
for determination of stellar parameters from oscillation frequencies.
They can be used to check how precise is \teff$ $ found by conventional methods.
{
If observational  \teff$ $ is precisely determined, then one can obtain metallicity 
from these relations.
The difference between two curves in {Fig. 8} for a given $\Delta n_{\rm x1} $ is about 250 K. The metallicity difference for the
models represented by these curves is 0.0150. This implies that  we can find $Z$
with an uncertainty of $\Delta Z=0.015$ if uncertainty in \teff$ $ is 250 K.
However, if $\Delta$\teff$ =100$ K, for example, $\Delta Z=0.006$.

For} 
some solar-like oscillating stars, only one minimum 
appears in the $\Dnu$-$\nu$ diagram.
In order to use asteroseismic relations for determination of fundamental 
stellar parameters, we must find out whether the seen minimum is min1 or min2 (see {Fig. 12}). 
%\textcolor{blue}{WHICH MINIMUN IS IT?  DO YOU MEAN MIN1 AND MIN2?  THIS
%NEEDS TO BE MORE CLEAR.}
If \teff$ $ and $Z$ of a given star are known, then we can 
overcome this problem again by using {equations (12) and (13). }
However, role of the convective parameter $\alpha$ must also be tested.

{The mass of the convective zone ($M_{\rm CZ}$) significantly 
depends on metallicity.}
%konkatM.vs.Dnu_max-min1DDnu.Z.gnu  in sisHbul
The generalized form of { $M_{\rm CZ}/\MS=0.066\Delta n_{\rm x1}$ (see fig. 6 of Paper I)} is 
\begin{equation}
\frac{M_{\rm CZ}}{\MS}=\left(0.225 Z+0.0027\right)\Delta n_{\rm x1}+0.627 Z-0.011
%\frac{M_{\rm CZ}}{\MS ?}=\left(0.00387\frac{Z}{{\rm Z}_{\sun}}+0.00267\right)\Delta n_{\rm x1}+0.0108\frac{Z}{{\rm Z}_{\sun}}-0.0110
\end{equation}
for arbitrary $Z$. 
$M_{\rm CZ}$ is directly proportional to $Z$, as expected.

% for determination of \teff$ $ ~ or for 
%determination of $Z$ 
%Diger nicelikler icin de (M,R,L,yas,g) Dnx1 cinsinden aciklamalar bulunabilir.
%Hic biri Teff ve Mcz gibi net bir iliskiye sahip degil.

%How the relations for L and R depend on Z?

\subsubsection{Effects of helium abundance}

{All of the models used in analysis of the metallicity effect 
are constructed with the helium abundance $Y=0.2804$.
In order to test influence of helium abundance on various asteroseismic 
relations, we also { obtain models } with $Y=0.2404$, $0.2604$, $0.2804$, 
$0.3004$ and $0.3204$.  
$T_{\rm eff}$-$\Delta n_{\rm x1}$ and $T_{\rm eff}$-$\Delta n_{\rm x2}$ 
relations do not depend on $Y$.
For stellar mass, we confirm that there is a moderate $Y$ dependence: }
% Teff.krs.Dnx1.gnu in  /home/yildiz/Kepler/Grids.2013/sisHbul
\begin{figure}
\includegraphics[width=100mm,angle=0]{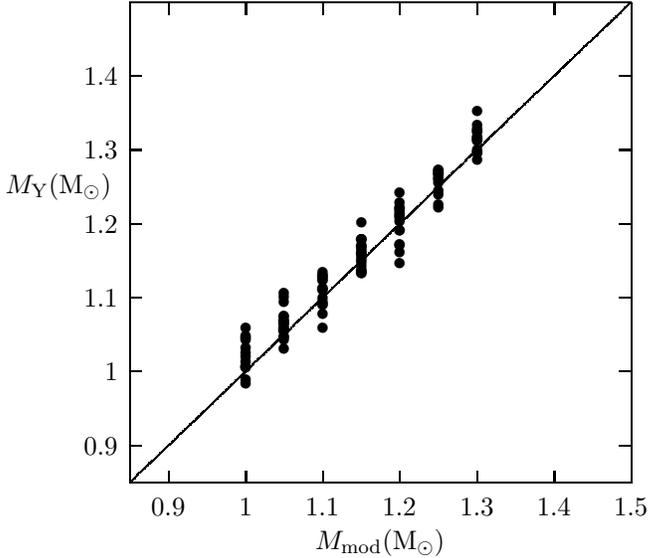}
%\includegraphics[width=100mm,angle=0]{M.krs.Msis.Xu1.pp.ps}
%\includegraphics[width=100mm,angle=0]{M.krs.Msis.Xu1.p.ps}
%480,width=140mm,angle=270]{the_four_diagrams16.ps}
\caption{ $M_{Y}$ given in {equation (15)} with respect to model mass
for the models with solar metallicity and $Y_0= 0.3204$, $0.3004$, $0.2804$, $0.2604$ and $0.2404$.
}
\end{figure}
\begin{equation}
\frac{M}{\MS}=\frac{\nu_{\rm min2}}{\nu_{\rm min2\sun}}\frac{\nu_{\rm max\sun}}{\nu_{\rm max}} 
\left( \frac{{{\rm Y}_{\sun}}}{Y}\right)^{0.25}.
\end{equation}
%$\beta_X=0.17$ for $\numin_2$ and $\beta_X=0.31$ for $\numin_1$.
%For better solution look for M/(sismik)= a(X)+b for each value of X
Mass computed from {equation (15)} is plotted with respect to model mass 
in {Fig. 9}.
 Although the data are scattered, the models are populated around the $M_Y=M_{\rm mod}$ line.
This implies that scaling relation also depends on helium abundance.
Thus, in order to find stellar mass and radius from asteroseismic quantities we also need 
{the} helium abundance. 
Therefore, we either { have to assume} that the helium abundance does not vary much from star to star
or find a new method for determination of the helium abundance (see Section 4.1).}
%The agreement between these masses is good enough. 
%Uncertainty is higher than the uncertainty
%in mass derived for the solar composition. 

If we combine {equations (8) and (15)}, then the seismic mass is 
proportional to $(Z/{\rm Z}_{\sun})^{0.12}(Y/{\rm Y}_{\sun})^{0.25}$. 
{Note especially that the power of $Y$ is twice the 
power of $Z$. }
The effect of metallicity on stellar mass is in general more important 
than that of $Y$, because range of $Z/{\rm Z}_{\sun}$ is much greater 
than $Y/{\rm Y}_{\sun}$.

{Equation (5)} for radius is not sensitive to $Y$ and therefore it can also 
be used for models with any helium abundance different from ${\rm Y}_{\sun}$.

\section{Determination of chemical composition from oscillation frequencies}
{The sound speed in a stellar interior is a function of 
the first adiabatic exponent $\Gamma_1$, pressure ($P$) and density ($\rho$): 
$c_{\rm }=\sqrt{\Gamma_1 P/\rho}$.  }
The local values of $\Gamma_1$ in the He ionization zones, however, depend on 
the He (or H) abundance and influence the spacing of oscillation frequencies. 
{The amplitude of the oscillatory component in the spacing
is determined by the He abundance (Houdek \& Gough 2011).  }
Similar correlation can be sought for metallicity. 
In the present section, we plot ${\Delta \nu-\braket{\Delta \nu}}$ with respect 
to $n$ for interior models with different chemical composition and try to find 
relations between amplitudes and chemical abundances.  
%\textcolor{blue}{WHAT IS N GRAPH?  THIS SHOULD BE MORE CLEAR.}

\subsection{Determination of helium abundance from oscillation frequencies}
\begin{figure}
\includegraphics[width=164mm,angle=0]{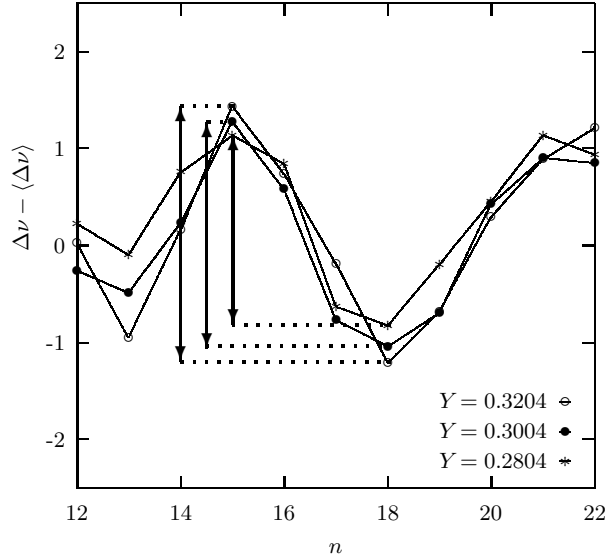}
%480,width=140mm,angle=270]{the_four_diagrams16.ps}
\caption{ $\Dnu-\braket{\Dnu}$ of 1.0 \MS models with ${ X_{\rm c}=0.17}$ and different initial helium abundances 
($Y_{\rm}= 0.2804$, $0.3004$ and $0.3204$) is plotted with respect to $n$. 
 The vertical arrows represent {the amplitudes $A_Y$.}
We notice that amplitude depends on $Y$. 
It increases as helium abundance increases. The highest one (the longest arrow) is for $Y_{\rm}= 0.3204$. 
{ The asterisks,} filled circles and  open circles are for $Y_{\rm}= 0.2804$, $0.3004$ and $0.3204$, respectively.
}
\end{figure}
\begin{figure}
\includegraphics[width=100mm,angle=0]{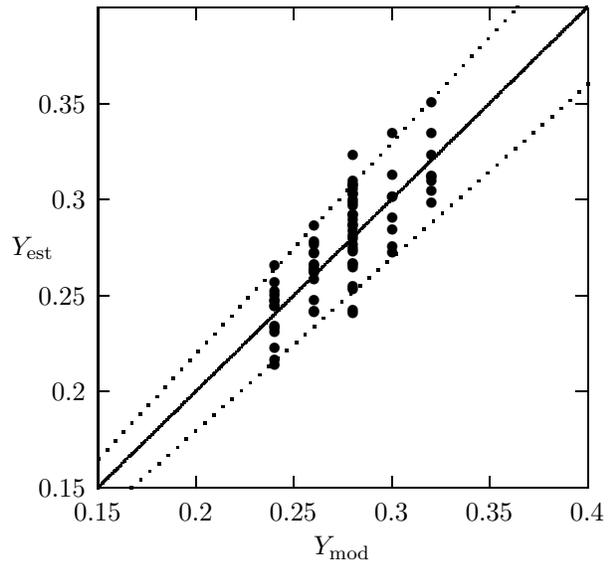}
%480,width=140mm,angle=270]{the_four_diagrams16.ps}
\caption{
Estimated $Y_{\rm est}$ by using {equation (16)} with respect to model $Y_{\rm mod}$.
The solid line is for $Y_{\rm est}=Y_{\rm mod}$. The upper and lower dotted lines are for $Y_{\rm est}=1.1Y_{\rm mod}$
and $Y_{\rm est}=0.9Y_{\rm mod}$, respectively.
}
\end{figure}

Determination of helium abundance from the second difference of oscillation 
frequencies is extensively discussed in several  papers 
(see, e.g. Miglio et al. 2010; Mazumdar et al. 2014). 
%Altthough helioseismology is very succesfull in determination of solar He abundance in the He {\scriptsize II} ionization zone
%(see, for example, Dziembowski, Pamyatnykh \&  Sienkiewicz 1991;
%Vorontsov, Baturin \& Pamyatnykh 1991; Perez Hernandez
%\& Christensen-Dalsgaard 1994; Basu \& Antia 199)
%{ Give a short summary of methods and results from literature}.
%
{Due to the effect of the He {\scriptsize II} ionization zone 
on the sound speed, helium (or hydrogen) abundance influences the 
spacing between the oscillation frequencies. }
We plot ${\Delta \nu-\braket{\Delta \nu}}$ with respect to $n$ for the interior models with the same 
input parameters but helium abundance.
In general, two maxima and two minima are seen in 
such a graph.
We already call the minima with the higher frequency (or order) as min1 
and the lower one as min2.
In a similar manner, we define max1 and max2 (see {Fig. 12}).
We confirm that the amplitude between min1 and max2  ($A_Y$) is a function of 
helium abundance. 
This is shown in {Fig. 10}, in which ${\Delta \nu-\braket{\Delta \nu}}$ 
is plotted with respect to order $n$. 
The frequencies are from three 1.0 \MS models with 
$Z_{\rm}=0.0172$ {and $X_{\rm c}=0.17$}, but different helium abundances. 
Their helium abundances are $Y_{\rm}=0.2804$, $0.3004$ and $0.3204$.
%The difference between the three models is in their 
%initial hydrogen abundance: $X_{\rm}=0.6624$, $0.6824$ and $0.7024$. 
%The amplitude between min1 and max2 is a fuction of $Y_{\rm}$. 
The highest amplitude occurs for the model with $Y_{\rm}=0.3204$. 
$A_{Y}$ gradually increases as $Y_{\rm}$ increases. 
This dependence has a diagnostic potential for determination 
of helium abundance. 
However, $A_{ Y}$ is not only a function of $Y$ but also \Dnu, $M$ 
(or \numin$_1/$\numax $~$) and $\braket{\delta \nu_{02}}$.
Using the models with $Y=0.2404$, $0.2604$, $0.2804$, $0.3004$ 
and $0.3204$, and $M=$ 1-1.3 \MSbit, we derive an expression for $Y$ as 

\begin{equation}
Y_{}= 10.74\frac{A_Y}{\Dnu}\left(\frac{\MS}{M}\right)^{1.5}+0.0034\braket{\delta \nu_{02}}+0.071.
%Y_{}= 0.100 A_Y+0.075.
\end{equation}
The estimated $Y$ by using {equation (16)} is plotted \wrt model $Y$ in {Fig. 11}.
There is a good agreement between estimated and model $Y$ values. 
The maximum difference between the estimated and model helium abundances 
is about 10 per cent.
{Equation (16)} can be used to estimate helium abundance of $Kepler$ and 
$CoRoT$ target stars.

%In Fig. 13, initial helium abundance is plotted with respect to $A_{ Y}$.
%The data are from the models with 1.0 and 1.2 \MS models with $X_{c}=0.35$ and $0.17$.
%There is a linear relation between $Y$ and $A_Y$.
%If $A_Y$ is derived from observed frequencies, helium abundance can be estimated
%by applying pure asteroseismic methods:  
%\begin{equation}
%Y_{}= 10.74\frac{A_Y}{\Dnu}\left(\frac{\MS}{M}\right)^{1.5}+0.0034\braket{\delta \nu_{02}}+0.071
%%Y_{}= 0.100 A_Y+0.075.
%\end{equation}
%where $A_Y$ is in units of $\muHz$.
%The uncertainty in $Y$ is unfortunately not very low. 
%However, if applicable, equation (19) can be used to estimate helium abundance of Kepler and CoRoT stars.
%
Since helium abundance in the He {\scriptsize II} ionization zone moderately 
changes due to microscopic diffusion, {equation (16)} does not directly give 
us initial helium abundance if age is not very small. 
{The value we find is the present value of $Y$, and it can 
be used as a constraint during calibration of interior models.}
However, there is significant difference between $A_Y$ of the Sun and 
solar model. 
While $A_Y$ of the Sun is about 3 $\muHz$, it is about 2 $\muHz$ for 
the solar models. 
{The so-called near-surface effects may influence $A_Y$. 
If the oscillation frequencies are corrected by using the method given by Kjeldsen, Bedding \& 
Christensen-Dalsgaard (2008), we obtain $A_Y=2.3$ \muHz. 
Although the near-surface effects decrease the discrepancy between observed and model values,
the remaining part is still significant.
%Thus, the near-surface effects partly improve the problem.
%This difference may be a result of differences between model and observed 
%oscillation frequencies caused by the near-surface effects. 
Consideration of $Kepler$ and $CoRoT$ stars is required if this problem can 
be solved by a simple method based on calibration approach.
}

\subsection{Determination of metallicity from oscillation frequencies}
The effect of metallicity on the amplitude is much stronger if we take 
the amplitude between max1 and min2.
This fact is sketched in {Fig. 12} in which
%Metallicity is much more effective on ${\Delta \nu-\braket{\Delta \nu}}$ diagram than the helium abundance.  In Fig. 14, 
${\Delta \nu-\braket{\Delta \nu}}$ is plotted with respect to order $n$ for 1.25 \MS models {with ${ X_{\rm c}=0.35}$}
and different $Z_{\rm}$ values. 
{The first maximum (in the right part of {Fig. 12}) and the 
second minimum (in the left part) are significantly sensitive to metallicity. }
The difference between max1 and min2 ($A_Z$) varies very rapidly as $Z_{\rm}$ 
changes.
It is slightly less than 4 \muHz $ $ for $Z_{\rm}=0.0172$ and about 
2 \muHz $ $ for $Z_{\rm}=0.0222$. 
It is very small 
for $Z_{\rm}=0.0322$ (about 0.5 \muHz) and becomes negative for some 
higher values of $Z_{\rm}$.
The amplitude $A_Z$ is not only function of $Z$ but also $M$ 
(or $\numin_1/\numax$) and $\braket{\delta \nu_{02}}$:
%\begin{equation}
%A_Z= 1.345\frac{M}{\MS}(1+2.681\frac{Z_{\sun}}{Z})-0.141\braket{\delta_{02}}-2.105.
%\end{equation}
%
\begin{equation}
A_Z=3.606\frac{M}{\MS}\frac{Z_{\sun}}{Z}+1.345\frac{M}{\MS}-0.141\braket{\delta \nu_{02}}-2.105.
%(0.01*($8+3.)+(-0.00144*$11+0.0135))
\end{equation}

{From {equation (17)}, we take $Z/{Z_{\sun}}$ to the left-hand side 
and find an expression for Z in terms of asteroseismic quantities as}
\begin{equation}
\frac{Z}{Z_{\sun}}= \frac{3.606 (M/\MS)}{A_Z-1.345(M/\MS)+0.141\braket{\delta \nu_{02}}+2.105}.
%(0.01*($8+3.)+(-0.00144*$11+0.0135))
\end{equation}
%
%Expression for $Z$ as a function of $A_Z$ of Fig. 14 and $\braket{\delta \nu_{02}}$ is obtained as
%\begin{equation}
%Z_{}= 0.01A_Z-0.00144\braket{\delta \nu_{02}}+0.0435.
%%(0.01*($8+3.)+(-0.00144*$11+0.0135))
%\end{equation}
In {Fig. 13}, estimated metallicity given in {equation (18)} is plotted with respect to model metallicity.
The agreement between these two  metallicities is very good.
The maximum difference between the two metallicities is about 14 per cent.
 
{
For 1.0 \MS models with $Z_{\rm}> 0.0322$, the second minimum disappears. 
This implies that, at least for MS stars, if the second minimum of a star is not observed then either its 
mass is less than  0.9 \MS (see table 1 of Paper I) or its metallicity is higher than 0.0322.
}
\begin{figure}
\includegraphics[width=164mm,angle=0]{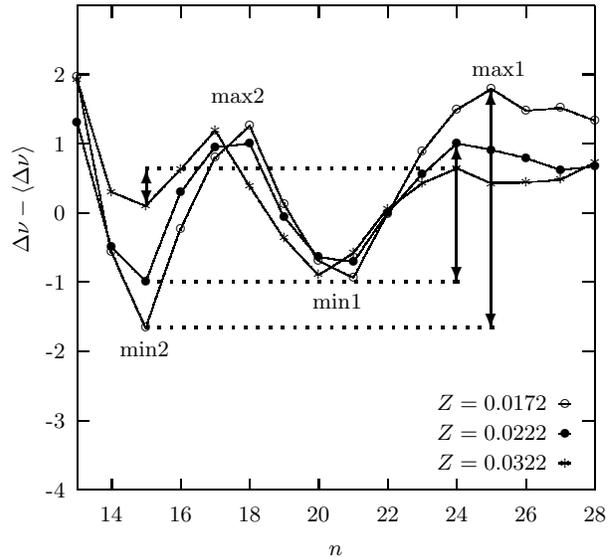}
%480,width=140mm,angle=270]{the_four_diagrams16.ps}
\caption{$\Dnu-\braket{\Dnu}$ of {1.25} \MS models with ${ X_{\rm c}=0.35}$ and different metallicities 
($Z_{\rm}= 0.0172$, $0.0222$ and $0.0322$) is plotted with respect to $n$.
{ The amplitude is defined by the vertical arrows.
It strongly depends on $Z$.
It decreases as metallicity increases. The amplitude is very small for the model with $Z_{\rm}= 0.0322$. }
}
\end{figure}
\begin{figure}
\includegraphics[width=100mm,angle=0]{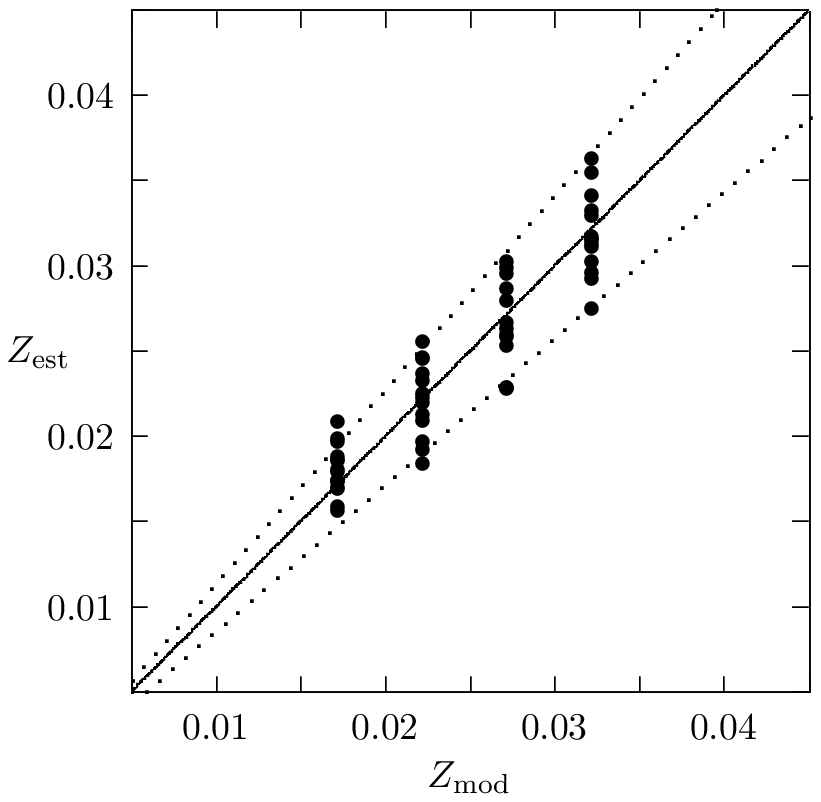}
\caption{Estimated $Z_{\rm est}$ by using {equation (18)} with respect to model $Z_{\rm mod}$.
The solid line is for $Z_{\rm est}=Z_{\rm mod}$. The upper and lower dotted lines are for $Z_{\rm est}=1.14Z_{\rm mod}$ 
and $Z_{\rm est}=0.86Z_{\rm mod}$, respectively.
}
\end{figure}

%\section{Fundamental properties of the Kepler and CoRoT stars from their oscillation frequencies}
%\subsection{Fundamental properties of the Kepler stars}
%The stars we study are the stars for which the oscillation frequencies are available in the literature.
%Apporchoux et al. (2012) give the freuencies and $\numax$ of 31 stars. We determine $\numin_1$ and $\numin_2$
%of these stars and then compute the masses of these stars by using different versions of asteroseismic expressions
%for the mass and other stellar properties. The results of our computations are given in Table 1.
% 
%\subsection{Fundamental properties of the CoRoT stars}

\section{Conclusion}
In Paper I, we have found new reference frequencies from the oscillation 
frequencies and derived new relations between asteroseismic quantities 
and all the fundamental stellar parameters.
{These relations are based on interior models for the 
mass range 0.8-1.3 \MS with the solar composition.}
In this study, the mass range is extended to 1.6 \MS and obtain new 
relations for arbitrary $Z$ and $Y$.

For the mass range $M>1.3$ \MSbit, the expressions given in Paper I are 
not valid any more and therefore we derive new relations for $M$, $R$, $L$, 
\teff, $M_{\rm CZ}$ and  $t$.

{Metallicity also affects stellar structure and evolution
significantly, and consequently affects oscillation frequencies. }
The derived relations are in general different for different $Z$.
We develop new relations valid for arbitrary $Z$.
{The relation between \teff,  $Z$ and $\Delta n_{\rm x1}$ 
({equation 12}) is in particular very useful and can be employed to determine 
of any of these quantities. }
We also obtain a similar relation for $\Delta n_{\rm x2}$ ({equation 13}).

The relations between asteroseismic and non-asteroseismic quantities are 
in general less sensitive to helium abundance in comparison with metallicity. 
However, the relation for mass is significantly changed by $Y$.
We find that estimated mass is inversely proportional to $Y^{0.25}$ 
({equation 15}).

We also develop new methods for determination of $Y$ and $Z$ from 
oscillation frequencies. 
We plot $\Delta \nu-\braket{\Dnu}$ with respect to $n$ for interior models with the same input parameters but
$Y$ or $Z$. 
These methods are based on the amplitudes in such diagrams.
Usefulness of these methods will be clear when they 
are applied to the $Kepler$ and $CoRoT$ targets.
The difference between model and estimated $Z$ ({equation 18}) values is 
about 14 per cent at most.
This difference is about 10 per cent for $Y$.

{
%We also analyze oscillatory components of \Dnu$ $ and show how the He {\scriptsize II} ionization zone
%determines the minima. According to our findings, deviation of frequencies from the asymptotic relation 
%is caused by difference between sound speed gradients at nodes around the He {\scriptsize II} ionization zone
%(see equation 6 and Fig. 2).

}

\section*{Acknowledgements}
Professor Chris Sneden  is acknowledged
for his suggestions which improved the presentation of the manuscript.
This work is supported by the Scientific and Technological Research Council of Turkey (T\"UB\.ITAK: 112T989).

\label{lastpage}

\end{document}